%% file: main.tex
\documentclass[useAMS,usenatbib]{mn2e}
\usepackage{graphicx}
\usepackage{amssymb, amsmath}
\usepackage{times}
\usepackage{color}

\usepackage{lineno}

\begin{document}


\title[BOSS RSD]{Testing General Relativity with Growth rate measurement from Sloan Digital Sky Survey III Baryon Oscillations Spectroscopic Survey galaxies}
\author[Alam et al.] {
    Shadab Alam$^{1,2}$, Shirley Ho$^{1,2}$,  
    Mariana Vargas-Maga\~na$^{1,2}$  , 
    Donald P. Schneider$^{3,4}$\\
    $^{1}$ Departments of Physics, Carnegie Mellon University, 5000 Forbes Ave., Pittsburgh, PA 15217 \\
    $^{2}$ McWilliams Center for Cosmology, Carnegie Mellon University, 5000 Forbes Ave., Pittsburgh, PA 15217 \\
    $^{3}$Department of Astronomy and Astrophysics, The Pennsylvania State University,University Park, PA 16802 \\
    $^{4}$ Institute for Gravitation and the Cosmos, The Pennsylvania State University, University Park, PA 16802}
    
\date{\today}
\pagerange{\pageref{firstpage}--\pageref{lastpage}}   \pubyear{2015}
\maketitle
\label{firstpage}

\input{tex/abstract}



\input{tex/intro}

\input{tex/theory}

\input{tex/data}

\input{tex/systematic}

\input{tex/analysis}

\input{tex/result}

\input{tex/discussion}



\bibliography{Master_Shadab}
\bibliographystyle{mnras}

%
\appendix
\input{tex/optimize}

\label{lastpage}

\end{document}

%% file: tex/abstract.tex
\begin{abstract}
The measured redshift ($z$) of an astronomical object is a combination of Hubble recession, gravitational redshift and peculiar velocity. The line of sight distance to a galaxy inferred from redshift is affected by the peculiar velocity component of galaxy redshift, which is observed as an anisotropy in the correlation function. This anisotropy allows us to measure the linear growth rate of matter ($f\sigma_8$).  We measure the $f\sigma_8$ at $z=0.57$ using the CMASS sample from Data Release 11 of Sloan Digital Sky Survey III (SDSS III) Baryon Oscillations Spectroscopic Survey (BOSS). The galaxy sample consists of 690,826 massive galaxies in the redshift range 0.43-0.7 covering 8498 deg$^2$. Here we report the first  simultaneous measurement of $f\sigma_8$ and background cosmological parameters using Convolution Lagrangian Perturbation Theory (CLPT) with Gaussian streaming model (GSRSD). We arrive at a constraint of $f\sigma_8=0.462\pm0.041$ (9\% accuracy) at effective redshift ($\bar{z}=0.57$) when we include Planck CMB likelihood while marginalizing over all other cosmological parameters. We also measure $b\sigma_8=1.19\pm0.03$, $H(z=0.57)=89.2\pm3.6$ km s$^{-1}$ Mpc$^{-1}$ and $D_A(z=0.57)=1401\pm23$ Mpc. Our analysis also improves the constraint on $\Omega_c h^2=0.1196\pm0.0009$  by a factor of 3 when compared to the Planck only measurement($\Omega_c h^2=0.1196 \pm 0.0031$). Our results are consistent with Planck $\Lambda$CDM-GR prediction and all other CMASS measurements, even though our theoretical models are fairly different. This consistency suggests that measurement of $f\sigma_8$ from Redshift space distortions at multiple redshifts will be a sensitive probe of  the theory of gravity that is largely model independent, allowing us to place model-independent constraints on alternative models of gravity.

\end{abstract}

%% file: tex/intro.tex
\section{Introduction}
\label{sec:intro}

The evolution of our Universe appears to be well described by the theory of general relativity(GR) \citep{Peebles1980,Davis83}. The predictions appear to be consistent with all the observations except the mysterious accelerated expansion of the Universe  \citep{Riess1998,Perlmutter1999}, and the dark matter \citep{Zwicky1937,Kahn1959,Rubin1970} . The accelerated expansion of the Universe proposed in $\Lambda$CDM-GR is in good agreement with Baryon Acoustic Oscillation (BAO) \citep{Eis2005,Cole2005,Hutsi2006,Kazin2010,Percival2010, Reid2010}, Hubble constant \citep{Riess2011} and Cosmic Microwave Background (CMB) \citep{Wmap2013, PlanckI}. Within the current paradigm, the primordial fluctuations in the early Universe were  amplified into structures we observe today via gravitational interactions. These gravitational interactions are the sum of the motions of matter and the expansion of the Universe. Therefore, one would only need to specify the initial conditions, the spatial geometry and the contents of the Universe to use Einstein's theory to predict the large-scale growth rate of the matter density in the Universe. We can compare such predictions to the observations in redshift surveys  \citep{BOSS, WiggleZ,6dFGRS,2dFGRS,Vipers}. The velocity field from maps of galaxies in such surveys can be measured because the galaxy redshifts, from which distances are inferred, include components from both the Hubble flow and peculiar velocities from the comoving motions of galaxies.  Such surveys thus reveal an anisotropic distribution of objects \citep{Cole1995,Peacock2001,Samushia2010}; the anisotropy in the clustering encodes information about the formation of structure and provides a sharp test of the theory.

In galaxy redshift surveys, the distortion produced in the two-point correlation function due to the peculiar velocity component in the galaxy redshift is known as ``Redshift Space Distorion (RSD)''. \citet{Kaiser87} first developed a formalism that describes redshift space power spectrum by modifying the linear theory of large scale structure. \citet{Hamilton92} extended this approach to the two-point correlation function in real space. The seminal work of \citet{Socco2004} describes a more general dispersion model, which improves the Kaiser linear model by including higher order terms. \citet{Socco2004} also invoked the concept of general streaming model which was first introduced by \citet{Davis83} and further developed by \citet{Fisher95}. A combination of the Lagrangian Perturbation Theory model by \citet{Socco2004} and the Gaussian Streaming model was used to measure the linear growth rate ($f\sigma_8$) of the Universe  in Baryon Oscillation Spectroscopic Survey \citep[BOSS;][]{BOSS}, which is part of Sloan Digital Sky Survey III \citep[SDSS-III;][]{Eisenstein2011},  DR9 \citep{ReiWhi11,Reid12} and DR11 \citep{Samushia13} data releases.  Other methods and models of the correlation functions or power spectrum are also used to derive $f\sigma_8$ with BOSS data \citep{Chuang13,Beutler13,Ariel13}. Other galaxy redshift surveys such as SDSS DR7 \cite{Cullan14},6-degree Field Galaxy Redshift Survey (6dFGRS,\cite{6dFGRS}), 2-degree Field Galaxy Redshift Survey (2dFGRS,\cite{2dFGRS}), WiggleZ \citep{WiggleZ}, and VIMOS Public Extragalactic Redshift Survey (VIPERS,\cite{Vipers}) have also measured redshift space distortion at different redshifts.

In this paper we employ a model called Convolution Lagrangian Perturbation Theory with Gaussian Streaming Redshift Space Distortions (hereafter CLPT-GSRSD)  to analyze BOSS DR11 \citep{And14}.
In order to test the model and the analysis method presented in this paper to a high accuracy, we have used a relatively large number of mock galaxy catalogs \citep{Manera13} with clustering properties similar to those of the higher redshift BOSS galaxies.  We also provide systematic errors based on the results of analyzing these mock galaxy catalogs. 

%% file: tex/theory.tex
\section{Theory}
\label{sec:theory}

In this section we describe the model with which we fit both the mock galaxy catalogs and BOSS data. 

Throughout the paper, we adopt the standard ``plane-parallel'' or ``distant-observer'' approximation, in which the line-of-sight direction to each object is taken to be the fixed direction $\hat{z}$. This approach has been shown to be a good approximation at the level of current observational error bars (e.g., Figure 10 of \citet{SamPerRac12} or Figure 8 of \citet{YooSel14}).

\subsection{CLPT} 

Convolution Lagrangian Perturbation Theory (CLPT) is a non-perturbative resummation of Lagrangian perturbation theory  \citep{Carlson12}. With CLPT, Carlson identified a few terms that asymptote to constants in the large-scale limit and hence need not be expanded with approximations. The authors showed that CLPT performs  better than all the other methods when compared to N-body simulations of dark matter halos. The monopole of correlation function matches N-body up to a very small scale; the quadrupole has less than a few percent error for scale above 20 h$^{-1}$Mpc (see Figure 1 and 2 of \cite{Carlson12}) . Unfortunately, the CLPT doesn't perform well in the quasi-linear regime for the quadrupole of biased tracer (see Figure 5 of \cite{Carlson12}). To overcome this problem with the quadrupole for a biased tracer we use CLPT in combination with the Gaussian Streaming Model (GSM) as it has been demonstrated to model the galaxy correlation functions more accurately \citep{Wang13}.
 
\subsection{The Gaussian streaming model}

The Gaussian Streaming Model (GSM) developed by \citet{ReiWhi11},  fits the monopole and quadrupole of the correlation functions of mock galaxies with a large-scale bias $b\simeq 2$ to the percent level on scales above $25\,h^{-1}$Mpc.  This model has been used to interpret the clustering of galaxies measured in BOSS \citep{Reid12,Samushia13,Samushia14}.

The GSM is inspired by the Eulerian streaming models.
It enforces pair conservation, assuming that the functional form of the halo velocity distribution is Gaussian, centered at $\mu v_{12}$, where $\mu v_{12}$ is mean line of sight velocity between a pair of tracers as a function of their real space separation.  Specifically we assume that the redshift-space halo correlation
function is
\begin{equation}
  1 + \xi^s(s_\perp, s_\parallel) =
  \int \dfrac{dy}{\sqrt{2\pi}\sigma_{12}}
  [1 + \xi(r)]
  \exp\left\{-\frac{[s_\parallel-y-\mu v_{12}]^2}{2(\sigma_{12}^2+\sigma_{FOG}^2)} \right\}\ ,
\label{eqn:gsm}
\end{equation}
where $\xi(r)$, $v_{12}$ and $\sigma_{12}$ are provided from an analytic theory.  In the model of \citet{ReiWhi11,Reid12} intergrated Lagrangian perturbation theory with scale-dependent but local Lagrangian bias \citep{Mat08b} was used for the real-space correlation function ($\xi(r)$) of halos, but the halo infall velocity ($v_{12}(r)$) and dispersion($\sigma_{12}(r)$) were computed in standard perturbation theory with scale-independent bias.
In order to move from halos to galaxies, \citet{Reid12} showed that it suffices
to introduce a single additional parameter, $\sigma_{FOG}$, akin to the $\sigma$ in Eq.~(\ref{eqn:gsm}).
This quantity  is taken to be an isotropic, scale-independent dispersion that is added in quadrature to $\sigma_{12}$  that modifies the scale-dependence of the quadrupole moment on small scales. There are more comprehensive simulation based models for describing the velocity distribution of galaxies around groups and clusters \citep{Zu13} and at large scales \citep{Bianchi15}, but it is relatively difficult to embed them into the halo model to explain the kinematics of the galaxies.

\subsection{CLPT-GSRSD} 

The quadrupole prediction on quasi-linear scales from CLPT for a biased tracer is not as accurate as its predictions for N-body simulations. 
CLPT was further improved by the model proposed by \citet{Wang13}, which combines the velocity statistics and correlation function from CLPT with GSM in order to produce  a more accurate monopole and
quadrupole for biased tracer. This model has  less than 5\% error in quadrupole for pair separation greater than 20 h$^{-1}$Mpc
for a biased tracer \citep{Wang13}. This model  is similar to the Zeldvoich Streaming Model (ZSM) \citep{White14} and Lagrangian Streaming Model (LSM) \citep{White15}. None of these models has been yet used to interpret the clustering of the BOSS galaxy sample. We will be using the analytical model described by \citet{Wang13} to extract constraints on cosmological parameters; we denote this model CLPT-GSRSD. 


%% file: tex/data.tex
\section{SDSS III- BOSS data} 
\label{sec:data}
\label{sec:boss} 

We use data included in data releases 10 (DR10;\cite{Ahn2014}) and 11 (DR11;\cite{Alam2014}) of the Sloan Digital Sky Survey (SDSS; \cite{York2000}). Together, SDSS I, II \citep{Abazajian2009} and III \citep{Eisenstein2011} used a drift-scanning mosaic CCD camera \citep{Gunn1998} to image over one-third of the sky (14555 square degrees) in five photometric bandpasses \citep{Fukugita1996,Smith2002,Doi2010} to a limiting magnitude of $r <22.5$ using the dedicated 2.5-m Sloan Telescope \citep{Gunn2006} located at the Apache Point Observatory in New Mexico. The imaging data were processed through a series of pipelines that perform astrometric calibration \citep{Pier2003}, photometric reduction \citep{Lupton1999}, and photometric calibration \citep{Padmanabhan2008}. All of the imaging was reprocessed as part of SDSS Data Release 8 (DR8; \cite{Aihara2011}).
BOSS \citep{Dawson2013}is designed to obtain spectra and redshifts for 1.35 million galaxies over a footprint covering 10,000 square degrees. These galaxies are selected from the SDSS DR8 imaging and are being observed together with 160,000 quasars and approximately 100,000 ancillary targets. The targets are assigned to tiles using a tiling algorithm that is adaptive to the density of targets on the sky \citep{Blanton2003}. Spectra are obtained using the double-armed BOSS spectrographs \citep{Smee2013}. Each observation is performed in a series of 900-second exposures, integrating until a minimum signal-to-noise ratio is achieved for the faint galaxy targets. This ensures a homogeneous data set with a high redshift completeness of more than 97\% over the full survey footprint. Redshifts are extracted from the spectra using the methods described in \citet{Bolton2012}. A summary of the survey design appears in \citet{Eisenstein2011}, and a full description is provided in \citet{Dawson2013}.

We use the CMASS sample of galaxies  \citep{Bolton2012} from data release 11 \citep{Alam2014}. The CMASS sample has 690,826  massive galaxies covering 8498 square degrees in the redshift range $0.43<z<0.70$, which correspond to an effective volume of 6 Gpc$^{3}$.

%% file: tex/systematic.tex

\section{Investigating the Systematics Budget}



  
In this section we will examine possible sources for the systematics. We will  first describe the result of fitting PTHalo mocks and then the systematic introduced by various approximations used in the analysis. We will conclude this section by describing the possible observational systematics.

\subsection{Mock Galaxy catalogs}
\label{sec:sim}

To validate CLPT-GSRSD we have used PTHalo mock catalogs.  Such validations have been performed by \citet{Wang13}  in the paper in which they were introduced. However, our goal is to test the model for mock galaxies with properties similar to those of the BOSS galaxies at $z\simeq 0.5$.  


The perturbation theory mock (PTHalo mock) is generated by populating a matter field using second order Lagrangian perturbation theory and calibrating the masses of dark matter halos by comparing it to detailed numerical simulations \citep{socco2001,Manera13}. We are using the PTHalo mocks generated and validated by \citet{Manera13} for DR11 footprint of SDSS-III survey. There are 600 PTHalo mocks available, which we employed to validate the model and decide upon some constraints on the model parameters. The simulations used to produce PTHalo mocks cover the same volume as that of the CMASS sample. These mocks are designed to have a bias similar to the bias of CMASS sample. It is important to note that both, mocks and our model,      are based on a similar kind of perturbation theory.

%
%
%
%
%
%

\subsection{Fitting PTHalo mocks}
\begin{figure}
\includegraphics[width=0.5\textwidth]{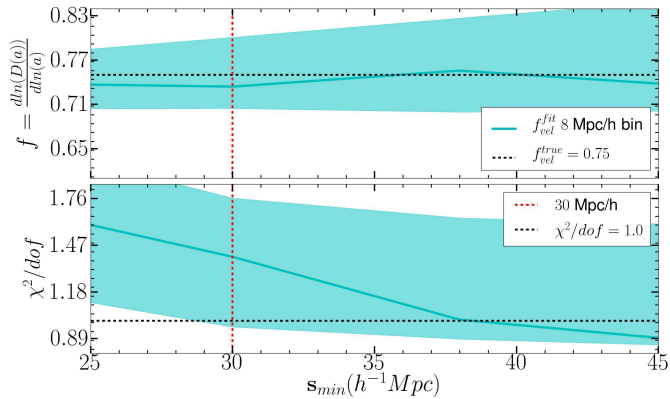}
\caption{The result of fitting 600 PTHalo mocks. The two panels show the mode and $1\sigma$ (computed from percentile ) spread of $\{f,\chi^2/dof \}$ as the function of $s_{min}$ minimum scale used for fitting the correlation function. In the top panel for $f$, we recover the expected value of the parameter within a few percent. The vertical dashed line indicates the  minimum scale used in this paper $s_{min}=30$ h$^{-1}$Mpc. The lower panel demonstrates that the model and data do not agree as we move to smaller(more non-linear) scale}
\label{fig:pthalofit}
\end{figure}

We have examined the result of fitting 600 PTHalo mocks. We fixed the cosmology while fitting these mocks because we precisely know the cosmology of simulation.  The mocks are constructed in such a way that they mimic the CMASS sample . We have computed the monopole and quadrupole of the correlation function for each of the 600 mocks using Landy-Szalay estimator \citep{LandySzalay93} with bins of 8 h$^{-1}$Mpc in pair separation. The monopole and quadrupole  are fit with the CLPT-GSRSD model for fixed fiducial cosmology and three freely-floating RSD parameters,  $\{F^\prime,F^{\prime \prime},f \}$, where the parameters represent the first and the second order Lagrangian bias and the derivative of logarithm of the growth factor, respectively. Figure \ref{fig:pthalofit} shows the result of fitting 600 PTHalo mocks using CLPT-GSRSD model as the function of minimum scale ($s_{min}$) used for the fit.  The top panel demonstrates that we recover the expected value of $f=d(\ln D)/d(\ln a)$ within a few percent for the entire range of $s_{min}$ shown in the figure. The bottom panel reveals that the $\chi^2/dof$ increases as we include smaller scales in the fit. The solid blue lines in these plots are the mode of the results from 600 mocks and the shaded region corresponds to $1\sigma$ limit. We have also performed this analysis with bins of 2 h$^{-1}$Mpc in monopole and quadrupole and found similar results. \citet{Wang13} has shown that the prediction of correlation function by CLPT-GSRSD agrees at few percent level down to $r=15$ h$^{-1}$Mpc with N-body simulation.   We would like to keep $s_{min}$ as small as possible because the correlation function has a high signal to noise at smaller scales. But, our mocks are based on a perturbation theory which means the shape of quadrupole might not be very accurate at small scale due to the non-linear velocity structure. It is clear from Figure \ref{fig:pthalofit} that below 36 h$^{-1}$Mpc the growth rate estimated starts to get slightly biased and $\chi^2/dof$ crosses 1. So, this is similar to bias-variance trade off, where $s_{min}=36$ h$^{-1}$ Mpc shows no bias and $s_{min}=22$ h$^{-1}$Mpc is with smaller variance. We have decided to use $s_{min}=30$ h$^{-1}$ Mpc as the optimum minimum scale with acceptable bias and low variance in our analysis. We have also repeated the analysis with $s_{min}=22$ h$^{-1}$Mpc and found consistent result with marginal improvement in precision.  

\begin{figure}
\includegraphics[width=0.5\textwidth]{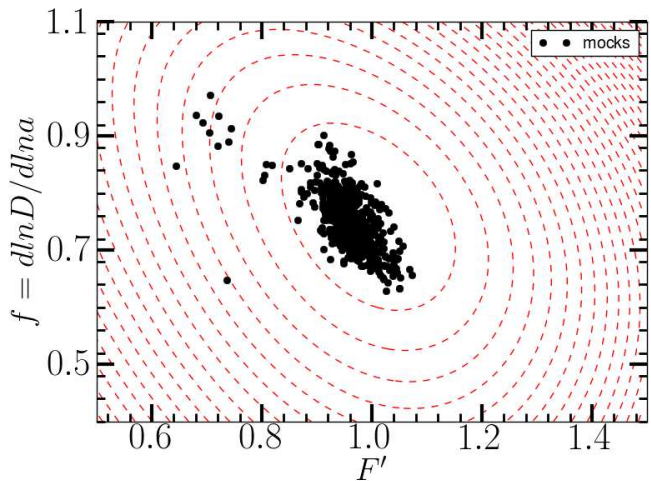}
\caption{The scatter shown by black points is the result of fitting 600 PTHalo mocks. Seven out of 600 mocks showed a best fit that is outside the region shown in the plot. The red-dashed contour lines represent the model degeneracy. This plot demonstrates that most of scattering in the mock can be explained by the model degeneracy. We have found 17 outlier mocks with $F\prime<0.8$ and relatively high $\chi^2$.}
\label{fig:degeneracy}
\end{figure}

Figure \ref{fig:degeneracy} demonstrates that the observed scatter in the parameter produced by fitting the mocks is completely consistent with the model degeneracy. The black points are the best fit obtained from the mock (some of the mocks are outside of the region shown in the plot). The red dashed contours in the figure are an estimate of the model degeneracy. We have evaluated the theoretical correlation function on the $100 \times 100$ grid in the region covered in Figure \ref{fig:degeneracy}. We have used correlation function evaluated for $F^\prime=1.0$ and $f=0.76$ as the reference model to evaluate the $\chi^2$ at each point of the grid using the covariance matrix evaluated from the mocks. The $\chi2$ surface constructed in this manner should reveal the degeneracy of the model. The red dashed line in Figure \ref{fig:degeneracy} represents the contour of $\chi^2$ while using fitting scale $30$ h$^{-1}$Mpc $ <s <$ 126 h$^{-1}$Mpc. We designate this as model  degeneracy and show that almost all the scatter in the mocks can be explained by the model degeneracy.  We have found 17 mocks with $F\prime<0.8$ (outliers). The mean $\chi^2/dof$ for the outlier mocks is 7.7, which is relatively high. The mean of quadrupole moment of these mocks shows stronger variation around $r=100$ h$^{-1}$Mpc.


\subsection{Observational effects}

There are several possible sources of observational systematics in the measurement of correlation function. These systematics have been studied in great detail for the SDSS DR9 and DR11 samples in \citet{ARoss12, And13,And14}. They have introduced two different systematic weights to reduce the cross correlation of $\xi_l$ with the star density and seeing, while demonstrating that potential systematics, such as sky brightness, do not affect the measurements. We have adopted the weights $w_{star}$ and $w_{see}$ in our measurement to remove these systematic effects. We have also used the weights to account for redshift failure ($w_{zf}$) and close pairs ($w_{cp}$). $w_{cp}$ accounts for the fact that the spectrum of only one of the galaxies is observed for galaxies separated by less than 62'' due to the finite size of fiber fittings. \citet{ARoss12} demonstrates that these weights don't change the cosmological result for Baryon Acoustic Oscillation (BAO) but do remove the systematic effects in the measurement of $\xi_l$ and improve the goodness of fit.  \citet{ARoss12} have also shown that $\xi_l$ for scales above $150$ h$^{-1}$Mpc has a systematic error larger than the statistical error. This issue is not a concern for our study because our maximum scale of interest is $126$ h$^{-1}$Mpc.

\begin{table}
\begin{center}
\caption{The shift in the parameter \{$f\sigma_8,\alpha_\parallel,\alpha_\perp$\} for each weight application from the default of applying all weights in units of error on the parameter. The stellar weight is most important for the full shape analysis.}
\label{tbl:sys}
\begin{tabular}{lccc}
Weights & $\Delta f\sigma8 / \sigma_{f\sigma_8}$ &
 $\Delta \alpha_{\parallel}/ \sigma_{\alpha_{\parallel}}$ & 
 $\Delta \alpha_{\perp}/ \sigma_{\alpha_{\perp}}$ \\
\hline
\hline
$w_{star}*(w_{cp} +w_{zf}-1)$   & $ -0.03$ &  0.02  &  0.03 \\
$w_{see}*(w_{cp} +w_{zf}-1)$    &  0.54  &  0.82  &  0.12 \\ 
$w_{cp} +w_{zf}-1$                    &  0.51  &  0.82  &  0.13 \\
$w_{zf}$                                     &  0.79  &  0.71  & -0.20 \\
$w_{cp}$                                    & 0.44   &  0.94  &  0.36 \\
None (uniform weighting)           & 0.72   &  0.82  &  0.06\\
\hline
\end{tabular}
\end{center}
\end{table}

We have also examined the effect of using different weighting on the measured RSD parameters (similar to \citet{Keisuke2014}) using observed correlation function. We have calculated the correlation function with different weighting, these correlation functions are fit for RSD and BAO parameters \{$f\sigma_8,bias,\sigma_{FOG},\alpha_\parallel, \alpha_\perp$\}. Table \ref{tbl:sys} lists the observed shift compared to the default case of using all weights. The stellar weight has the important effects in our measurements. If we do not use any weights the shifts are of the order of $1\sigma$ for $f\sigma_8$ and $\alpha_\parallel$ but is small for $\alpha_\perp$. The shifts in measurement due to these systematic weights will be higher as we will improve precision in future surveys.


%% file: tex/analysis.tex

\section{Analysis} 
\label{sec:analysis} 

In this section we describe the steps of our analysis starting from galaxy position to the parameter constraints. We first briefly mention the different steps then elaborate on the procedure with details in the following subsections . 

We first measure the galaxy correlation function using SDSS-DR11 (CMASS) galaxy sample. We have used 600 PTHalo mocks to generate an estimate of the covariance matrix. We start our optimization problem with nine-dimensional parameter space consisting of four cosmological parameters and three RSD parameters, and run Markov Chain Monte Carlo(MCMC) to explore this parameter space using COSMOMC \citep{cosmomc}. In every MCMC step, we first evaluate the Planck likelihood, and then evaluate the linear power spectrum for the current sampled(grid) point if it has not been evaluated previously. This linear power spectrum is fed to the CLPT theory to calculate the correlation function($\xi(r)$) and velocity statistics. The CLPT $\xi(r)$ and velocity statistics are used by the Gaussian streaming model GSRSD with the three RSD parameters to evaluate the redshift space two-dimensional correlation function. This two-dimensional correlation function is rescaled according to the difference in the fiducial cosmology and the current MCMC cosmology to determine the final model correlation function. The final correlation function is used as the theoretical model (CLPT-GSRSD) to calculate the $\chi^2_{RSD}$ with data correlation function and mock covariance matrix. The  $\chi^2_{RSD}$ is converted to a likelihood, which is multiplied to the Planck likelihood in order to calculate the total likelihood, which is maximized using COSMOMC.

The basic approach is to estimate the cosmological parameters by fitting the monopole and quadrupole of correlation function from SDSS CMASS DR11 using the correlation function of CLPT-GSRSD in combination with Planck likelihood computed from CMB power spectrum.

\subsection{Measuring the correlation function and covariance matrix}
We first assumed the fiducial cosmology as flat $\Lambda$CDM-GR cosmological model with $\Omega_m=0.274$, $H_0=0.7$ ,$\Omega_bh^2 =0.0224$, $n_s=0.95$ and $\sigma_8=0.8$  \citep{And14} in order to convert observed celestial coordinates ($\alpha,\delta$) and redshift to the position of the galaxy in three-dimensional space. These galaxy positions are used to estimate the two point statistic (correlation function) of the galaxy using the minimum variance estimator Landy-Szalay estimator \citep{LandySzalay93}.
\begin{equation}
\hat{\xi}(\Delta r) =\frac{DD(\Delta r) -2DR(\Delta r) +RR(\Delta r)}{RR(\Delta r)}
\end{equation}
where DD, DR and RR represent weighted pair count of galaxy, cross pair count of galaxy-random  and pair counts of randoms, respectively. We used the weighting $w_{tot}=w_{fkp} w_{star} w_{see} (w_{cp}+w_{zf}-1)$ as described in \citet{And14} to correct for observational systematic \textcolor{red}{errors} and minimize variance. The correlation function is function of $r$, which is the distance between a pair of galaxies and $\mu=cos(\theta)$, where $\theta$ is the angle between separation vector and line of sight. We compress the information by projecting the correlation function to the Legendre polynomial $L_l(\mu)$ of order $l$ as follows.

\begin{equation}
\hat{\xi}_l (r) = \frac{2l+1}{2} \int_{-1}^1 d\mu \hat{\xi}(r,\mu)L_l(\mu)
\approx \frac{2l+1}{2} \sum_{k} \Delta \mu_k \hat{\xi}(r,\mu) L_l(\mu_k)
\end{equation}

\citet{Wang13} reported that the CLPT-GSRSD model is a good fit to N-body simulation for $l=0$ and $l=2$ but not for higher $l$, therefore, we will limit our analysis up to $l=2$. The signal measured in the correlation function should evolve with redshift due to the evolution of $\sigma_8(z)$, but a good approximation is to use an effective redshift for DR11 sample $\bar z =0.57$ as shown in equation[10-12] of \citet{Samushia13}. 

To estimate the uncertainty in our measurement of the correlation function, we constructed the covariance matrix using 600 PTHalo mocks. The inverse of the sample covariance  estimated from finite number of mocks is a biased estimator. We will adopt the correction for covariance matrix as described in \citet{Maria13,Will14}. The sample covariance matrix is calculated as follows.

\begin{equation}
\hat{C}_{i,j}= \frac{ \sum_{m=1}^{300} (\xi_i-\bar \xi)(\xi_j -\bar \xi) +   \sum_{m=301}^{600} (\xi_i-\bar \xi)(\xi_j -\bar \xi)}{2\times299}
\end{equation}
where $\hat{C}_{i,j}$ represents the covariance between bin $i$ and $j$ and $\bar \xi$ is the mean of the mocks and the sum is over different mocks. The corrected inverse covariance matrix is

\begin{equation}
C^{-1}_{i,j}= (1-0.62(2 \times n_{r_{bin}} +1)/(N_{mock}-1) )\hat{C}^{-1}_{i,j}
\end{equation}
where $N_{mock}$ is the total number of mocks which is 600 for us, and $n_{r_{bin}}$ is the number of bins of the correlation function used in the analysis. 

\subsection{Parameter space}
Our model parameters can be subdivided into two subsets. The first set is of cosmological parameter ($\Omega_b h^2$ ,$ \Omega_c h^2$, $n_s$,$A_s$ $H_0$), where $\Omega_b$, $\Omega_c$ and $H_0$ are baryon density, dark matter density and hubble constant, respectively, with $h$ being $H_0/100$. The quantity $A_s$ is the scalar amplitude of primordial power spectrum. This choice of parameters requires us to assume an evolution model in order to evaluate the theoretical model at the galaxy redshift.We want our measurement to be independent of such an assumption, which is necessary to be able to use these results to test various models of gravity. This goal can be accomplished by fixing $H_0$ to the best fit and allowing two extra parameters, $H(z)$ and $D_A(z)$, at the effective redshift. This can be modeled using Alcock-Paczynski parameters $\alpha_\parallel, \alpha_\perp$. The second set of parameters are the redshift space distortion (RSD) parameters ($F^\prime$,$F^{\prime\prime}$,$f$,$\sigma_{FOG}$), where $F^\prime$,$F^{\prime\prime}$ and $f$ represent the first and the second order Lagrangian bias, and the logarithmic derivative of the growth factor, respectively. The parameter $\sigma_{FOG}$ is an additional isotropic velocity dispersion to account for the finger-of-god effect \citep{ReiWhi11}. The second order Lagrangian bias is not well constrained because it is more important for small scale quadrupole. We have considered two cases. In the first one we have marginalized over $F^{\prime\prime}$ with hard prior covering $[-5,5]$ in the second case we have sampled overdensity $\nu$ which determines both $F^\prime$ and $F^{\prime\prime}$ using peak background split \citep{White14}. 
\begin{align}
F^\prime &= \frac{1}{\delta_c} \left[ a\nu^2-1 +\frac{2p}{1+(a\nu^2)^p}\right] \\
F^{\prime\prime}&= \frac{1}{\delta_c^2} \left[ a^2\nu^4-3a\nu^2 +\frac{2p(2a\nu^2+2p-1)}{1+(a\nu^2)^p} \right]
\end{align}
where a = 0.707, p = 0.3 gives the Sheth-Tormen mass function \citep{ST1999}, and $\delta_c=1.686$ is the critical density for collapse. Independent of whether we use peak background split or marginalize over $F^{\prime\prime}$, we obtain consistent results .

\subsection{Calculating Theoretical model}
We first calculate the linear matter power spectrum using Code for Anisotropies in the Microwave Background (CAMB) \citep{Camb}. This approach requires the knowledge of cosmological parameters of the model we are evaluating. The linear power spectrum is then sampled according to the sampling scheme for $k$ described in Appendix \ref{sec:optimize}. The sampled power spectra is used to calculate the real space correlation function ($\xi(r)$), pairwise velocity statistics ($v_{12}(r)$) and the dispersion of pairwise velocity statistics ( $\sigma_{12}(r)$ ) as described in \citet{Wang13} and \citet{Carlson12}.

\begin{align}
1+\xi(r) &= \int d^3q M_o(r,q) \\
v_{12}(r) &= [1 + \xi(r)]^{-1} \int d^3q M_{1,n}(r,q) \\
\sigma_{12,nm}^2 (r) &= [1 + \xi(r)]^{-1} \int d^3q M_{2,nm}(r,q) \\
\sigma_\parallel^2 (r)  &= \sum_{nm} \sigma_{12,nm}^2 \hat{r}_n \hat{r}_m \\
\sigma_\perp^2 (r) &= \sum_{nm} (\sigma_{12,nm}^2 \delta_{nm}^K -\sigma_\parallel^2)/2
\end{align}
where $M_o(r,q)$ ,$M_{1,n}(r,q)$ and $M_{2,nm}(r,q)$ are integrals of the CLPT perturbation theory that depend on the linear matter power spectra. The $\hat{r}_n$ ,$\hat{r}_m$ are unit vectors along the galaxies position vectors. Please refer to \citet{Wang13} for details of these integrals and derivation of these equations.

The real space correlation function and velocity statistics calculated from CLPT together with the RSD parameters are supplied to the Gaussian Streaming Model (GSM) \citep{ReiWhi11} in order to evaluate the redshift space correlation function as follows.

\begin{multline}
1+\xi^{model}(s_\perp,s_\parallel) =  \\ 
\int \frac{dy [1+\xi(r)]}{[2\pi \sigma_{12}^2(r,\mu)]^{1/2}}\times exp\left\{ -\frac{[s_\parallel-r_\parallel-\mu v_{12}(r)]^2}{2\sigma_{12}^2 (r,\mu)} \right\} 
\end{multline}

\begin{equation}
\sigma_{12}^2(r,\mu) =\mu^2 \sigma_\parallel^2(r) +(1-\mu^2) \sigma_\perp^2(r) +\sigma_{FOG}^2
\end{equation}
where $s_\perp$ is the transverse separation in both real and redshift space. The quantities $s_\parallel$ and $r_\parallel$ are the LOS (line of sight) separation in redshift and real space, respectively. 

We want our measurement to be independent of any particular model of gravity. This can be achieved by avoiding the use of any model for the evolution of structure formation under the assumption of small deviation from widely accepted $\Lambda$CDM-GR. We will model this deviation by using two parameters ($\alpha_\parallel, \alpha_\perp$) which are defined as follows. 

\begin{equation}
\alpha_\parallel = \frac{H_{fiducial}}{H(z_{eff})} ,\alpha_\perp= \frac{D_A(z_{eff})}{D_A^{fiducial}} 
\end{equation}

We will rescale the model redshift space correlation function to account for this extra distortion as follow.

\begin{multline}
\xi_l^{RSD}(s) = \\
\frac{\displaystyle   \sum_{\lvert s-s_o \rvert <\Delta s/2} (2l+1)\xi^{model}(\alpha_\parallel s_\parallel, \alpha_\perp s_\perp) P_l(\mu) \sqrt{1-\mu^2}}{\frac{2}{\pi}\text{Number of bins used in sum}} 
\end{multline}
where $\{H_{fiducial},H\}$ and $\{D_A^{fiducial},D_A\}$ are the hubble expansion rate and angular diameter distance for the fiducial and model cosmology, respectively, $s_o=\sqrt{\alpha_\parallel^2 s_\parallel^2+\alpha_\perp^2 s_\perp^2}$ and $\Delta s =5$ h$^{-1}$Mpc. The above rescaling is simply the application of Alcock-Paczynski effect \citep{AP79}.

\subsection{MCMC and Likelihood estimation}
We use COSMOMC to perform Markov Chain Monte-Carlo likelihood analysis \citep{cosmomc} and explore nine-dimensional parameter space. We have four cosmological parameters $\{ \Omega_b h^2, \Omega_c h^2, n_s, A_s \}$. These parameters have flat prior centered at best fit value of Planck with width $\pm 5\sigma_{planck}$ (see Table \ref{tbl:parameter}). The other set of parameters are five redshift space distortion parameters $\{\nu,f,\sigma_{FOG},\alpha_\parallel,\alpha_\perp \}$, where  $\nu$ is overdensity, which determines the first and the second order bias using peak background split relation. When we marginalize over second order Lagrangian bias $F^{\prime\prime}$ in place of using the peak background split, then our second set of parameters is replaced by $\{F^\prime, F^{\prime\prime}, f,\sigma_{FOG},\alpha_\parallel,\alpha_\perp \}$.

We calculate the model monopole and quadrupole of the galaxy correlation function as described in the previous section for each point in the parameter space visited by the MCMC sampler. We have optimized our model evaluations with some assumptions, discussed in the Appendix \ref{sec:optimize}, in order to make time per MCMC iteration smaller. The likelihood constraint from RSD is calculated using $\chi^2$ as follows.

\begin{align}
\chi^2_{RSD}& =  ( \xi_{model} - \xi_{data})^T C^{-1} (\xi_{model} - \xi_{data} ) \\
\mathcal{L}_{RSD} &= \exp({-\chi^2/2})
\end{align} 
 where $\xi_{model}= [ \xi_0^{RSD}; \xi_2^{RSD}]$ , $\xi_{data}= [ \xi_0^{data}; \xi_2^{data}]$ and $C^{-1}$ is the inverse of corrected covariance matrix calculated from 600 PTHalo mocks. The cosmological parameters are well constrained from Planck satellite CMB data, which are mostly independent of the gravity and growth of structure parameters. Therefore, we will multiply our RSD likelihood by the Planck likelihood to obtain the joint constraint on our parameters.

\begin{equation}
\mathcal{L}_{total} = \mathcal{L}_{planck} \mathcal{L}_{RSD}
\end{equation}

The Planck likelihood is calculated using the constraint reported from planck temperature anisotropy data alone \citep{planck13}. The Planck parameter covariance is obtained from the correlation matrix given in Figure 21 of \citep{planck15}. We employ a multivariate Gaussian approximation to the full Planck likelihood. This approximation is close to the actual likelihood in the parameter space in which we are working.

\begin{equation*}
\Omega_b h^2 =0.02207 ,
\Omega_c h^2 =0.1196 ,
n_s =0.9616 ,
A_s =3.098
\end{equation*}

\begin{equation*}
\setlength{\arraycolsep}{0.5pt}
C_{planck} =
\scriptscriptstyle
\left(
\begin{smallmatrix}
1.089 \times 10^{-7}  & -4.501 \times 10^{-7} & 1.365 \times 10^{-6} &  3.564 \times 10^{-6} \\
-4.501 \times 10^{-7} &  9.610 \times 10^{-6} & -2.215 \times 10^{-5}  & 1.562 \times 10^{-5} \\
 1.365 \times 10^{-6}  & -2.215 \times 10^{-5} &   8.836 \times 10^{-5}   &  2.030 \times 10^{-5} \\
3.564 \times 10^{-6}  & 1.562 \times 10^{-5}   & 2.030 \times 10^{-5}  & 5.184 \times 10^{-3}
\end{smallmatrix}
\right)
\end{equation*}

%% file: tex/result.tex
\section{Results} 
\label{sec:results} 

\begin{table*}
\begin{center}
\caption{The list of parameters used in our analysis. For each parameter we provide their symbol, prior range, central value and $1\sigma$ error. We report the results of using peak background split to relate the first and second order bias and also the result of using sampling of the two bias independently. The results in both case are consistent. We also list the result when Planck prior is replaced by WMAP prior, which predicts $2\%$ shift in $f\sigma_8$.}
\label{tbl:parameter}
\begin{tabular}{lcccc}
Parameter    & prior range & \multicolumn{2}{c}{\textbf{Peak background split}} & \textbf{First and Second order bias}\\
 & &  with WMAP prior & with Planck prior  & with Planck prior \\
\hline
\textbf{Sampling Parameters} \\
\hline
$\Omega_b h^2 \dots$    & [0.02042 , 0.02372 ]& $0.02267\pm0.00036$& $0.02206\pm 0.00026$  & $0.02206 \pm 0.00026$\\
$\Omega_c h^2 \dots$    & [0.1041 ,    0.1351]  & $0.1141\pm0.0021$ & $0.11956 \pm 0.00086$  & $0.11956 \pm 0.00086$\\
$n_s \dots$                      & [0.914 ,  1.008]       & $0.9741\pm0.0085$& $0.9614 \pm 0.0058$      & $0.9613  \pm 0.0058 $\\
$\ln(10^{10}A_s) \dots$   & [2.67 ,   3.535]       &  $3.178\pm0.029$&  $3.093 \pm 0.066$        &  $3.103    \pm  0.070$\\
$\alpha_\parallel$             & [0.8 ,1.2]                &  $1.003\pm0.039$&  $1.051 \pm 0.043$         & $1.058 \pm 0.047$\\
$\alpha_\perp$                & [0.8 ,1.2]                 &  $0.997\pm0.018$&   $1.03   \pm 0.016$        & $1.032 \pm 0.016$\\
$f=dlnD/dlna$                  & [0.3 , 1.2]                &  $0.739\pm0.067$&  $0.747  \pm 0.072$        &  $0.729 \pm 0.073$\\
$\sigma_{FOG}$             &[0 , 10]                     &   $2.26\pm1.46$&  $1.91\pm 1.28$             &  $2.70 \pm1.69$\\
$v_{RSD}$ & [1.5 , 2.0] &$1.83\pm0.038$ &  $1.80 \pm 0.05$            &    \\
$F^{\prime}$                    & [0.5 , 1.5]                &    &                                         &$0.93 \pm 0.07$\\
$F^{\prime \prime}$                    & [0.5 , 1.5]                &    &                                         &$1.0 \pm 1.78$\\
\hline
\textbf{Derived Parameters} \\
\hline
$f\sigma_8$   & \dots                                    &  $0.454\pm0.041$&  $0.462 \pm 0.041$            & $0.453 \pm 0.041$ \\
$b\sigma_8$  & \dots                                    &  $1.21\pm0.030$ &  $1.194 \pm 0.032$             & $1.20 \pm 0.032$\\
$D_A( z=0.57)$ & \dots                                 &  $1356.0\pm24.0$ & $1400.9 \pm 22.7$             & $1403 \pm 21.9$\\
$H( z=0.57)$    & \dots                                  &  $93.4\pm3.6$ &  $89.2 \pm 3.6$                   & $88.5 \pm 3.9$\\
$F_{AP}$    & \dots                                        &  $0.663\pm0.033$ & $0.654 \pm 0.033$             & $0.651 \pm 0.034$\\
$D_V(z=0.57)$ & \dots                                  &  $2024.5\pm27.2$ & $2101.4 \pm 25.6$            &  $2108 \pm 29.4$\\
\end{tabular}
\end{center}
\end{table*}

\begin{figure}
\includegraphics[width=0.4\textwidth]{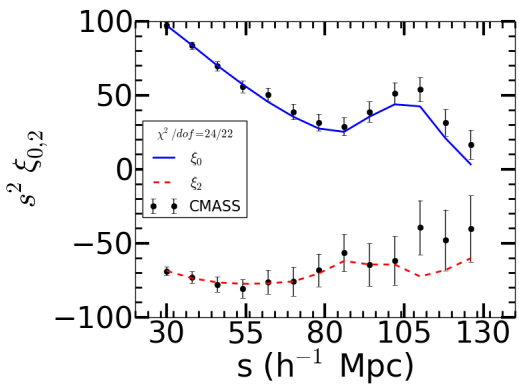}
\caption{This figure shows the projection of two-dimensional correlation function in legendre basis. The black data points are measured correlation function for DR11 CMASS sample and the error bars are the diagonal terms of covariance matrix calculated using 600 PTHALO mocks. The blue and red line represents the best fit monopole and quadrupole with $\chi^2/dof=24/22$.}
\label{fig:xifit}
\end{figure}

\begin{figure}
\includegraphics[width=0.20\textwidth]{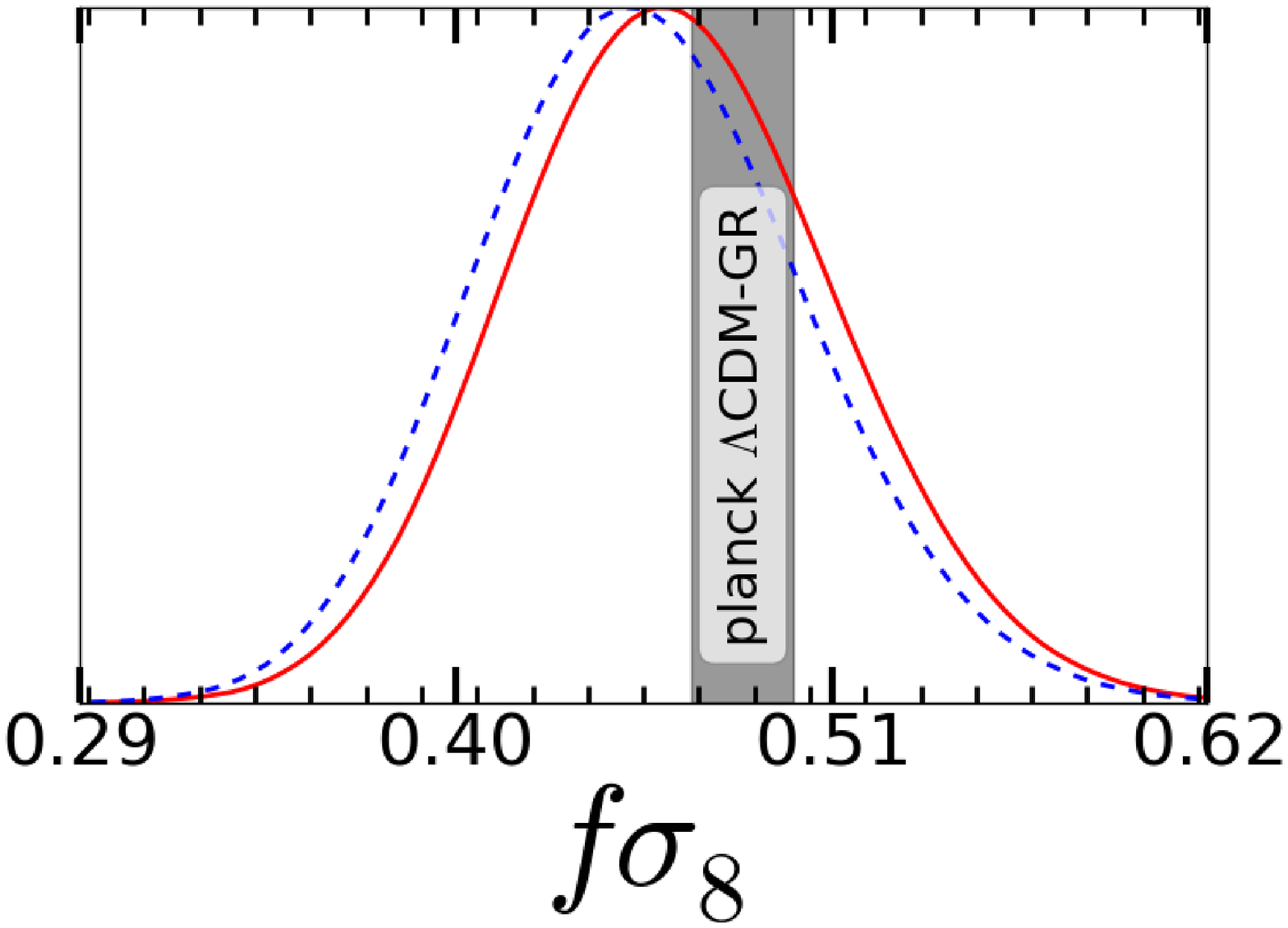}
\includegraphics[width=0.20\textwidth]{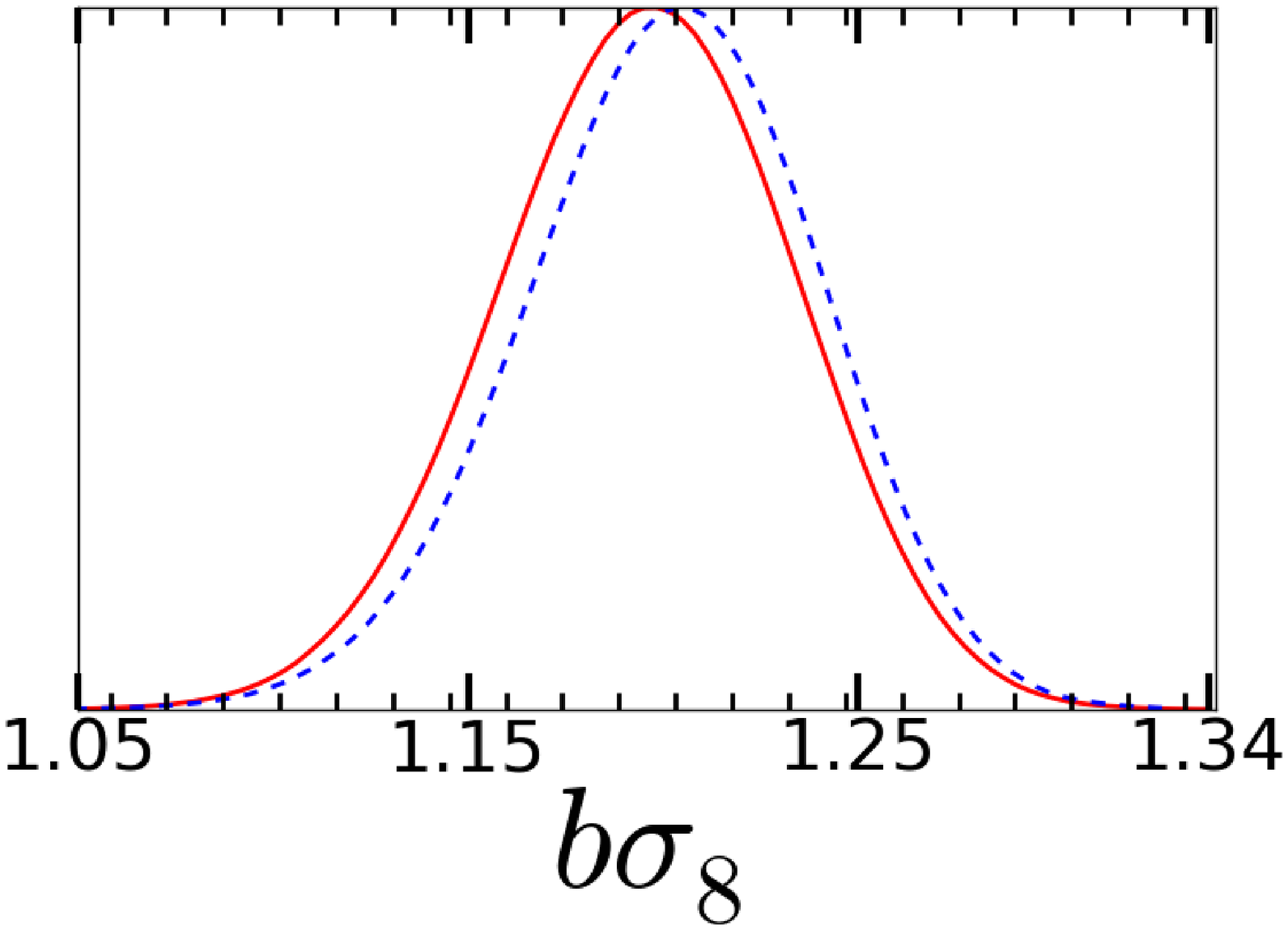} \\
\includegraphics[width=0.20\textwidth]{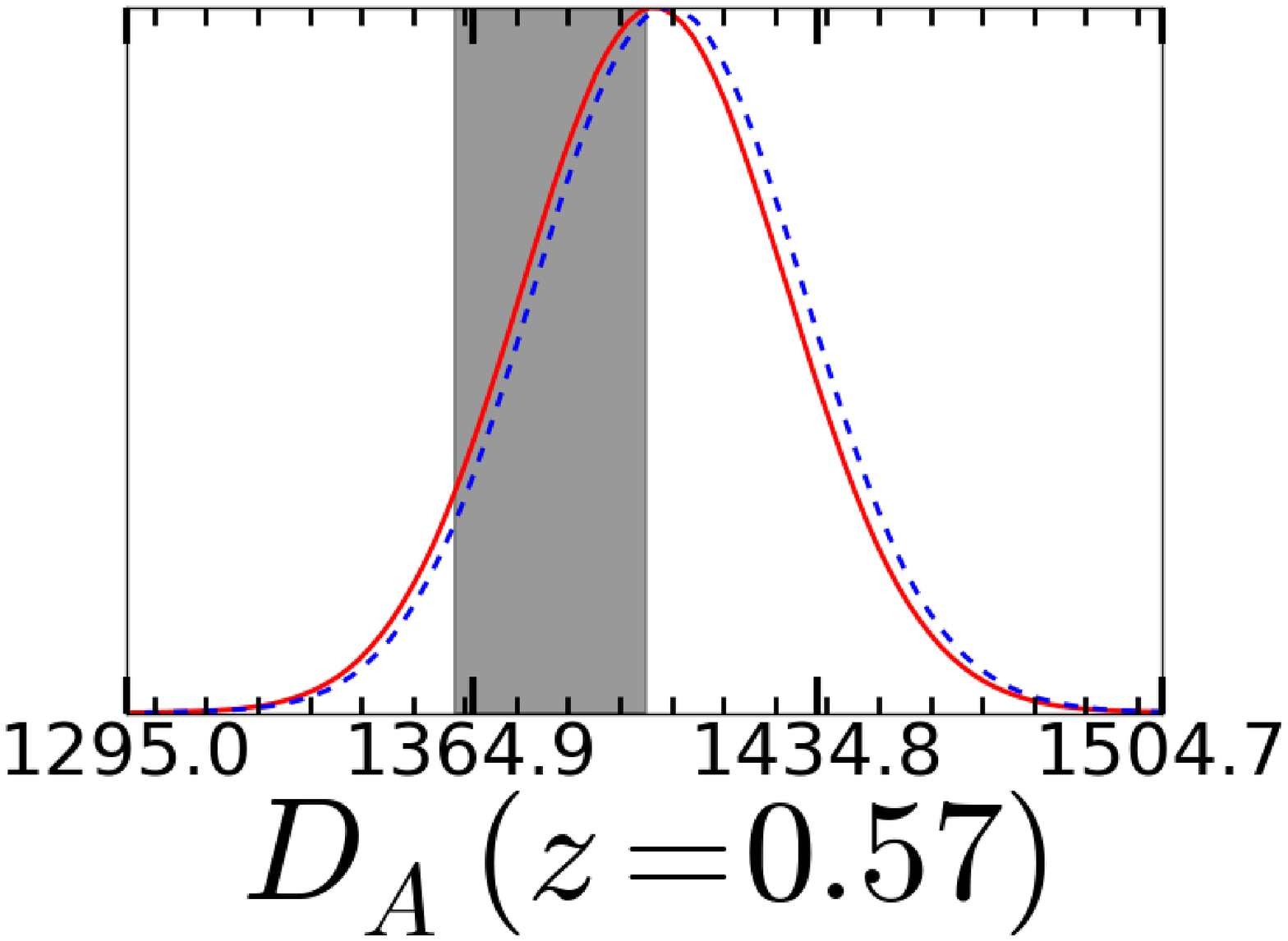}
\includegraphics[width=0.20\textwidth]{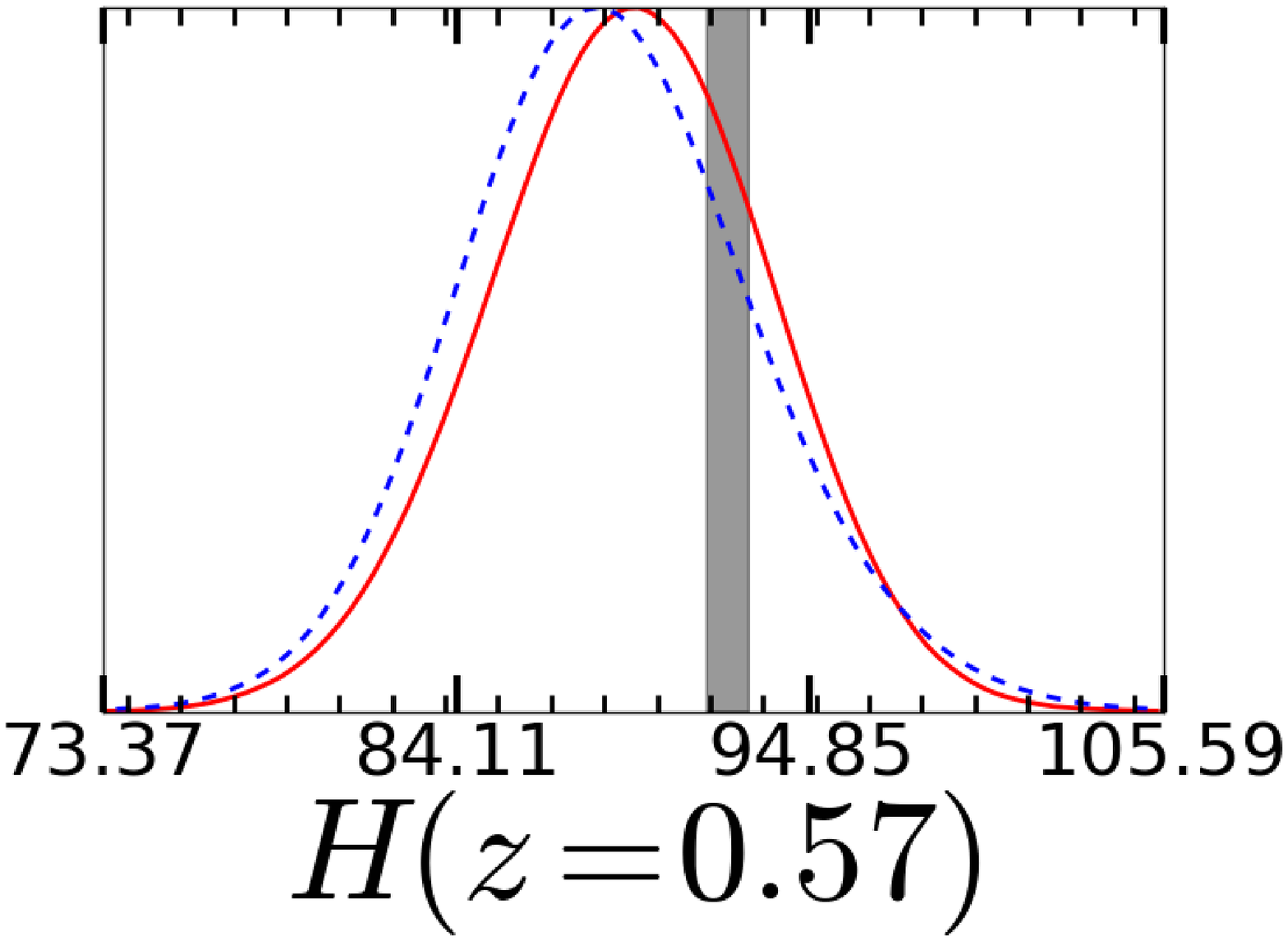}\\
\includegraphics[width=0.2\textwidth]{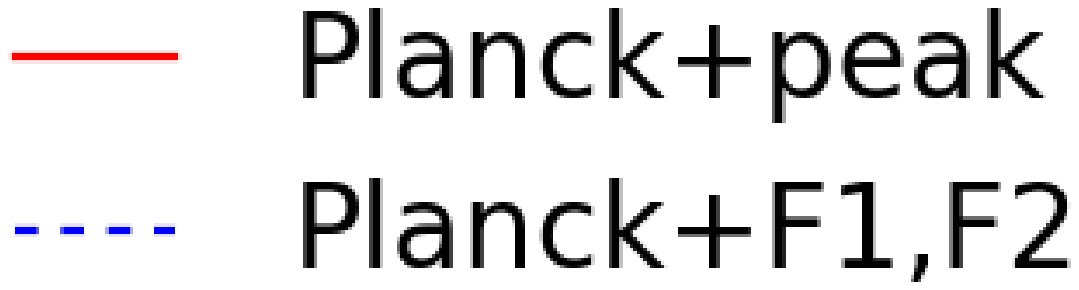}
\caption{The four panels show the one-dimensional marginalized likelihood for the parameters \{$f\sigma_8,b\sigma_8,D_A, H$\} at effective redshift 0.57 . The red solid line shows the result with peak background split, and the blue dashed line is the result when we fit for both first and second order Lagrangian bias. The grey shaded region shows 1$\sigma$ constraint from Planck with $\Lambda$CDM-GR. We detect 1.9\% shift in $f\sigma_8$ and less than 1\% for other parameters between the analysis with and without peak background split.}
\label{fig:RSD1d}
\end{figure} 

\begin{figure}
\includegraphics[width=0.22\textwidth]{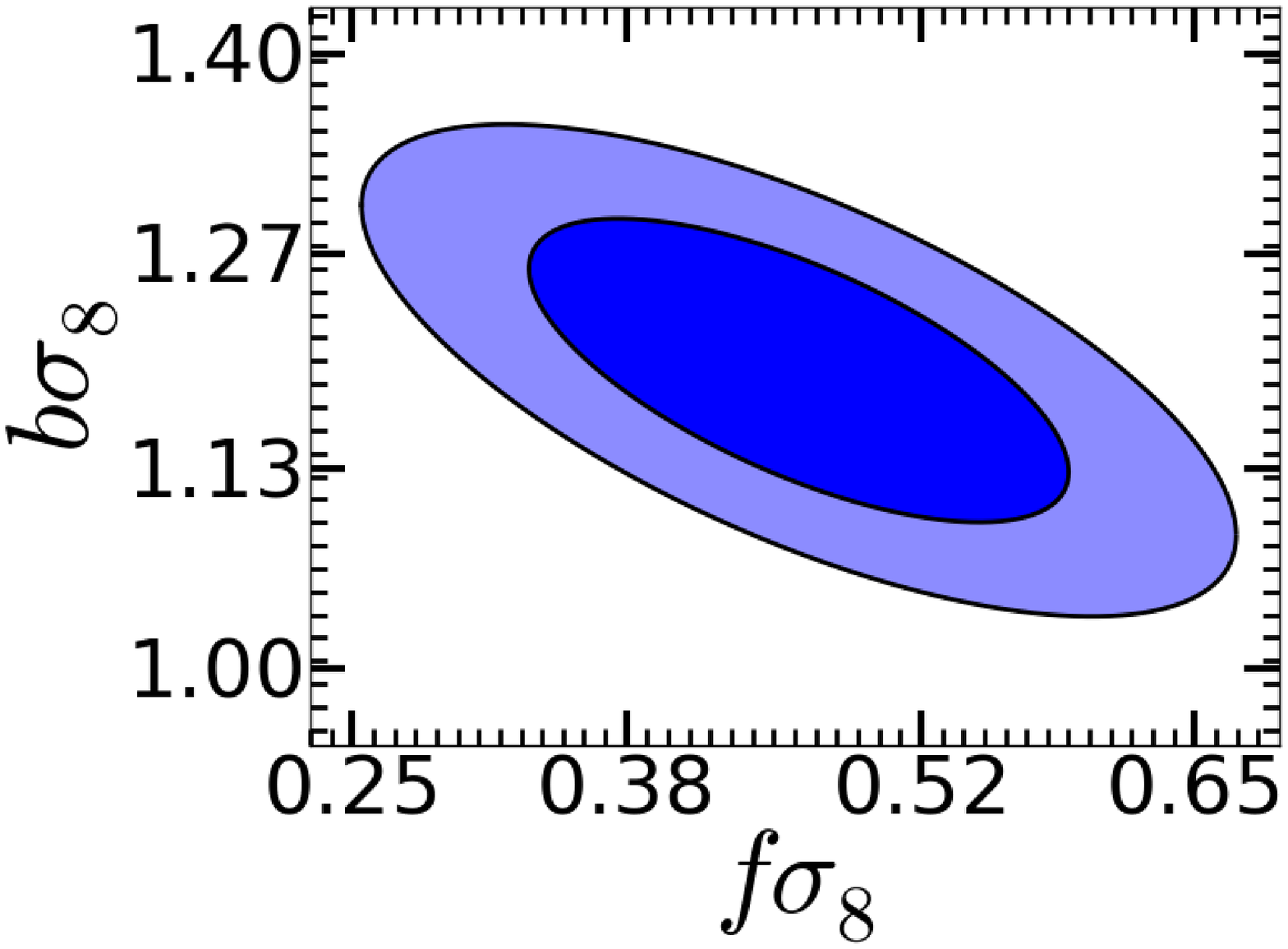}
\includegraphics[width=0.22\textwidth]{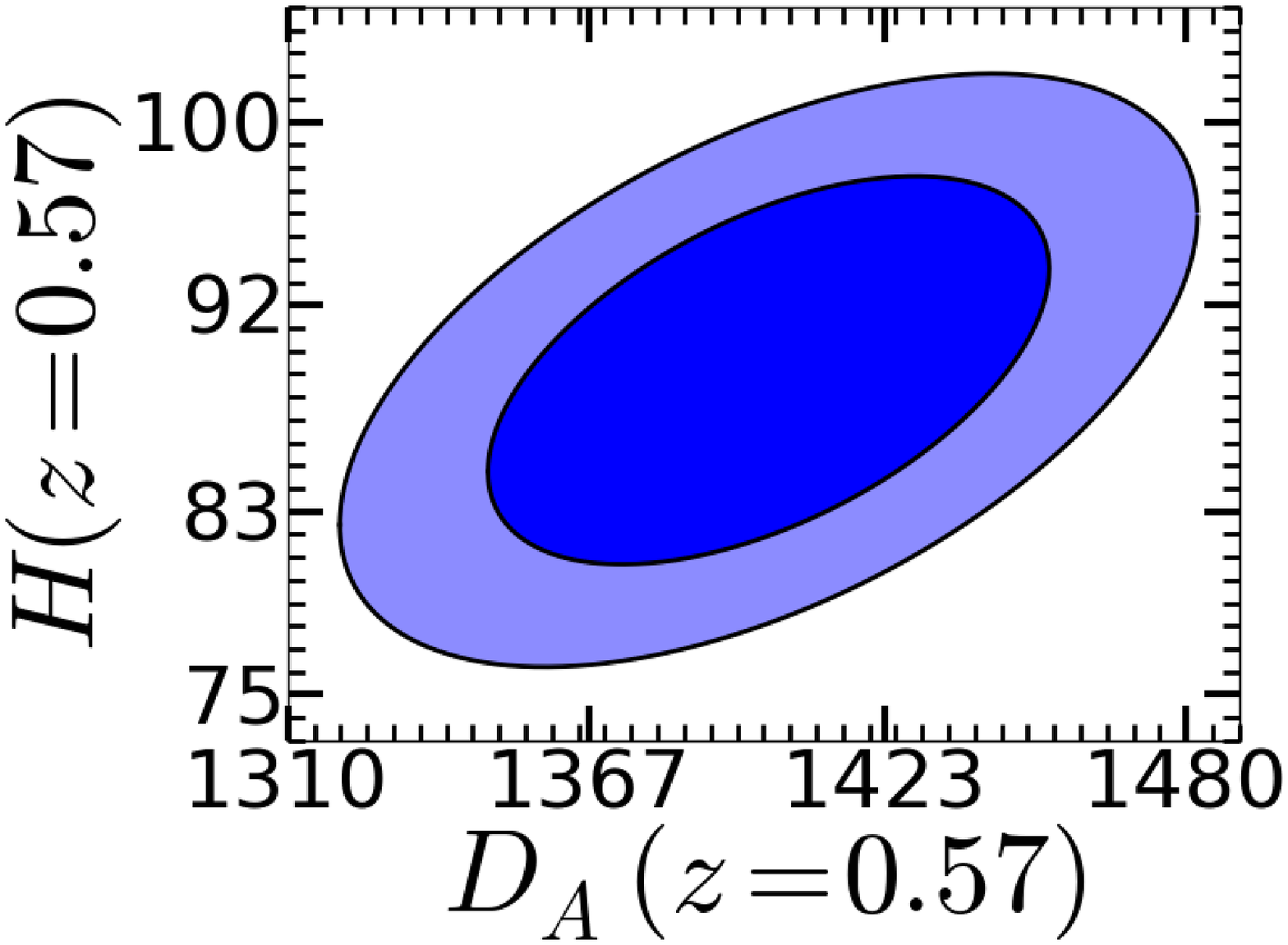}
\caption{Two-dimensional 68\% ($1\sigma$) and 95\% ($2\sigma$) confidence limits obtained on $f\sigma_8$---$b\sigma_8$ and $D_A$---$H$ at effective redshift of 0.57 recovered from Planck CMB and CMASS ($\xi_{0,2}$) datasets with peak background split assumption.}
\label{fig:Like2d}
\end{figure}

Figure \ref{fig:xifit} presents the monopole and quadrupole of the galaxy correlation function. The black data points are measurements from BOSS CMASS DR11 galaxy sample. The error bars on the measurements are the diagonal elements of mock covariance matrix. The blue and red line represents the best fit monopole and quadrupole, using the fitting range $30$ h$^{-1}$Mpc $ \leq s \leq 126 $ h$^{-1}$Mpc with $8$ h$^{-1}$Mpc sampling. The best fit $\chi^2/dof=24/22$ is achieved after marginalizing over essentially all the relevant parameters as listed in Table \ref{tbl:parameter}. The Figures \ref{fig:RSD1d} and \ref{fig:Like2d} show the one dimensional and two dimensional marginalized likelihood for some of the parameters. The final results of this analysis are given in Table \ref{tbl:parameter}. We have measured $f\sigma_8(z=0.57)= 0.462\pm0.041$ , $b\sigma_8 = 1.19\pm0.03$, $D_A(z=0.57) = 1401\pm23$ Mpc and $H(z=0.57)= 89.2\pm3.6$ km s$^{-1}$ Mpc$^{-1}$. The galaxy correlation function doesn't improve the constraints on the baryon density ($\Omega_b h^2$), scalar spectral index($n_s$) and amplitude of primordial curvature perturbation ($A_s$) at $k_0=0.05$ Mpc$^{-1}$h over the already tight constraints from Planck, however we do improve the measurement of cold dark matter  density ($\Omega_c h^2 = 0.1196\pm0.0009$) ,which is a $0.7\%$ measurement, this is an improvement in measurement of cold dark matter density compared to Planck ($\Omega_c h^2 = 0.1196\pm0.0031$) by a factor of 3.6 . We have also studied the impact of second order bias on growth rate measurement. We found consistent constraint while allowing both of first and second order Lagrangian bias to be free (shown by blue dashed line in Figure \ref{fig:RSD1d}). It is interesting to note that we have found  covariance between second order Lagrangian bias ($F^{\prime \prime}$) and growth rate. This suggest that a better modeling of higher-order bias including local and non-local contributions (for a recent work along this line, see e.g. \citet{Saito2014}) will become more important as the clustering statistics become more precise with future surveys.We have also repeated our analysis by replacing the Planck prior with WMAP prior \citep{Wmap2013} and found $2\%$ shift in $f\sigma_8$, which is much smaller than the estimated error.

\begin{figure}
\includegraphics[width=0.4\textwidth]{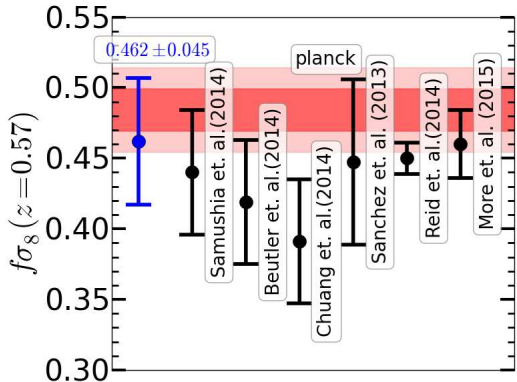}
\caption{Comparison of $f\sigma_8$ with other analysis on the same DR11 CMASS sample. The blue point present the result from our analysis. Our measurement is consistent with other clustering analysis and Planck $\Lambda$CDM-GR prediction. }
\label{fig:fsigma8}
\end{figure}

Our measurement of $f\sigma_8$ is consistent with all the other measurements reported from the same data set, as shown in Figure \ref{fig:fsigma8}. \cite{Samushia13} has reported $f\sigma_8(z=0.57)=0.44\pm0.044$, which is similar to our analysis. One major difference is in the theoretical model used in the two studies. We have used CLPT-GSRSD to evaluate our model correlation function, whereas \citet{Samushia13} use Lagrangian perturbation theory (LPT) as the model to predict correlation function. \citet{Beutler13} used the monopole and quadrupole of power spectrum and reported $f\sigma_8(z=0.57)=0.419\pm0.044$, whereas \citet{Chuang13} performs the analysis in configuration space with a different fitting model and obtained $f\sigma_8=0.391\pm0.044$. \citet{Ariel13} used wedges to measure the RSD signal and reported $f=0.719\pm0.094$. \citet{Reid14} has done the analysis at small scale $0.8$ h$^{-1}$Mpc $<s<32$ h$^{-1}$Mpc with halo occupation distribution model and Planck best fit cosmology and measured $f\sigma_8(z=0.57)=0.45\pm0.011$. \citet{Reid14} provides the strongest constraint on the growth rate but this analysis has significant modeling and cosmological assumptions. \citet{More2014} measured the constraint on $\Omega_m$ and $\sigma_8$ using a combination of abundance, clustering and galaxy-galaxy lensing. They have reported a constraint on $f\sigma_8$ by assuming the General Relativity linear theory prediction for growth rate ($f=\Omega_m^{0.545}$). Our measurement is competitive with all RSD measurements from large scale. 

%

%% file: tex/discussion.tex
\section{Discussion}
\label{sec:discussion}
We have presented an analysis of Redshift Space Distortion (RSD) using the SDSS-III BOSS DR11 CMASS sample, and have measured the monopole and quadrupole moments of galaxy auto correlation function at effective redshift of 0.57. We have used CLPT-GSRSD to model the Legendre moments of redshift space galaxy auto correlation function. The model used here does not work at small scales due to non-linearity, and measurement of correlation function from data shows systematic error at large scales. Therefore, we have adopted a conservative fitting scale between 30 h$^{-1}$Mpc and 126 h$^{-1}$Mpc, which we chose with the aid of a suite of perturbation theory mocks. Our measurements of linear growth rate ($f\sigma_8$), angular diameter distance ($D_A$) and Hubble constant ($H$) at effective redshift of 0.57 don't assume $\Lambda$CDM-GR evolution by virtue of using Alcock-Paczynski parameters ($\alpha_\parallel, \alpha_\perp$) independent of cosmology at current epoch ($z=0$). This approach makes these measurements suitable to test the predictions of various alternate models of gravity and cosmology.

Our results are consistent with \citet{Samushia13}, who performed a similar analysis on the same data set. However, the perturbation theory models used in the two analyses are different. Our model (CLPT) performed better on N-body simulation compared to the Lagrangian Perturbation Theory (LPT) model used in \citet{Samushia13}, . We have seen marginal improvement in the measurement uncertainty compared to the previous analyses. This is the first use of CLPT-GSRSD to measure both cosmology and growth simultaneously from the galaxy redshift survey. It has been used by \citet{Cullan14} to measure the growth rate with fixed cosmology for SDSS main galaxy sample. We couldn't use our model at smaller scales because our mocks cannot be trusted in this range. In the future we may be able to extend this model to scales as low as 20 h$^{-1}$Mpc if a reliable technique to test them on realistic mocks can be developed.

The linear growth factor has been measured in many redshift surveys between redshift of 0 and 1. Our measurement provides an important data point to study the evolution of the linear growth factor with redshift. The absolute value of $f\sigma_8$ and its evolution with redshift is quite sensitive to the model of gravity. These measurements will provide a good test of the general theory of relativity and the standard model of cosmology on the largest distance and time scales. It is possible to use these measurements to constrain flatness of the Universe and the dark energy equation of state parameter. These measurements also have the ability to constrain the parameters of alternate theories of gravity and dark energy. 

The next-generation surveys are going to be even more powerful, which will provide better measurement of correlation function and measurement of $f\sigma_8$, hence better understanding of cosmology and gravity. The error in the measurement of the correlation function is much smaller at small scales, which has not yet been explored in this paper due to our inability to test the theoretical model in this range. We can tap into the potential of small-scale clustering using RSD measurement when we can model the nonlinear clustering at small scale either analytically or using fast simulations.

\vspace{0.2in}
We would like to thank Lile Wang, Martin White and Beth Reid for providing the CLPT-GSRSD code. We also thank Keisuke Osumi for providing systematic weighted correlation functions as well Ross O' Connell for useful discussion. We like to thank Eric Linder and Martin White for their suggestions.
This work made extensive use of the NASA Astrophysics Data System and of the {\tt astro-ph} preprint archive at {\tt arXiv.org}.
The analysis made use of the computing resources of the National Energy
Research Scientific Computing Center.  This work is partially supported by NASA NNH12ZDA001N- EUCLID and  NSF AST1412966. S.H. and S.A. are partially supported by DOE-ASC, NASA and the NSF. 
Funding for SDSS-III has been provided by the Alfred P. Sloan Foundation, the Participating Institutions, the National Science Foundation, and the U.S. Department of Energy Office of Science. The SDSS-III web site is http://www.sdss3.org/.

SDSS-III is managed by the Astrophysical Research Consortium for the Participating Institutions of the SDSS-III Collaboration including the University of Arizona, the Brazilian Participation Group, Brookhaven National Laboratory, Carnegie Mellon University, University of Florida, the French Participation Group, the German Participation Group, Harvard University, the Instituto de Astrofisica de Canarias, the Michigan State/Notre Dame/JINA Participation Group, Johns Hopkins University, Lawrence Berkeley National Laboratory, Max Planck Institute for Astrophysics, Max Planck Institute for Extraterrestrial Physics, New Mexico State University, New York University, Ohio State University, Pennsylvania State University, University of Portsmouth, Princeton University, the Spanish Participation Group, University of Tokyo, University of Utah, Vanderbilt University, University of Virginia, University of Washington, and Yale University.

%% file: tex/optimize.tex
\section{Improving the efficiency of MCMC}
\label{sec:optimize}
The most computationally expensive part of likelihood analysis is the calculation of CLPT correlation function and velocity statistics. In order to make this high dimensional optimization problem reasonably efficient,  we have adopted two modifications. First we optimized the sampling of $k$ in the linear power spectrum used as the input to the CLPT perturbation theory. Second , we have discretized a small subspace of the parameter required for the perturbation theory in order to avoid doing almost the same calculation thousands of times.

 \subsection{Power Spectrum Sampling}
 The CLPT's runtime depends on the number of k points sampled in linear power spectrum. However, if we reduce the sampling of power spectrum too much, the integrals involved in the CLPT theory might not converge. Therefore, we need to minimize the sampling in $k$ to reduce the calculation time but keep the sampling sufficiently high to avoid the convergence problem. We have run an optimization of the sampling in $k$ by checking the convergence of the correlation function produced by CLPT and achieved the best case runtime of about 1 min for CLPT correlation function, which doesn't have any convergence issues in the scale of interest, which is up to 130 h$^{-1}$Mpc. 
 
\begin{figure}
\includegraphics[width=0.4\textwidth]{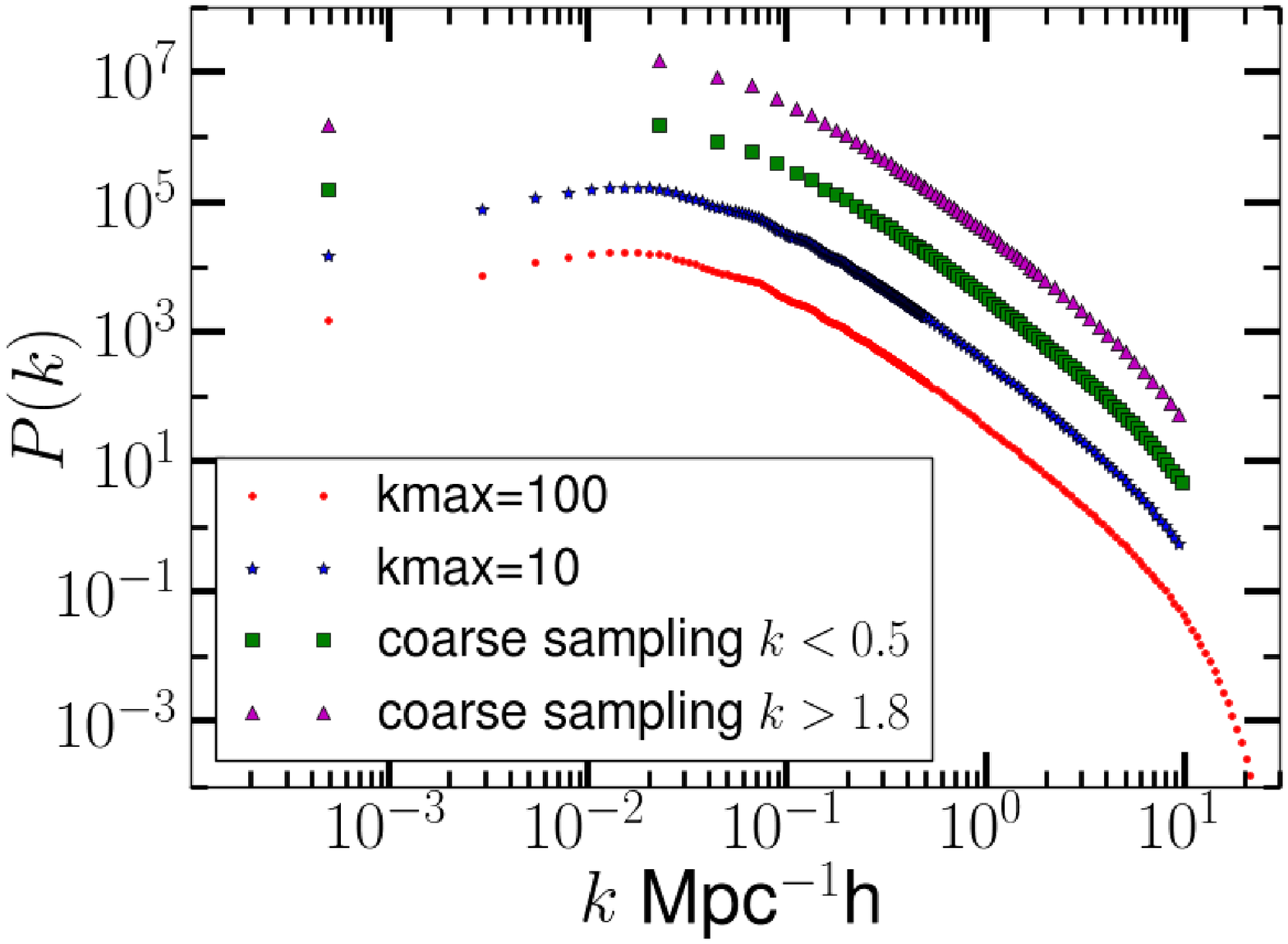}
\includegraphics[width=0.4\textwidth]{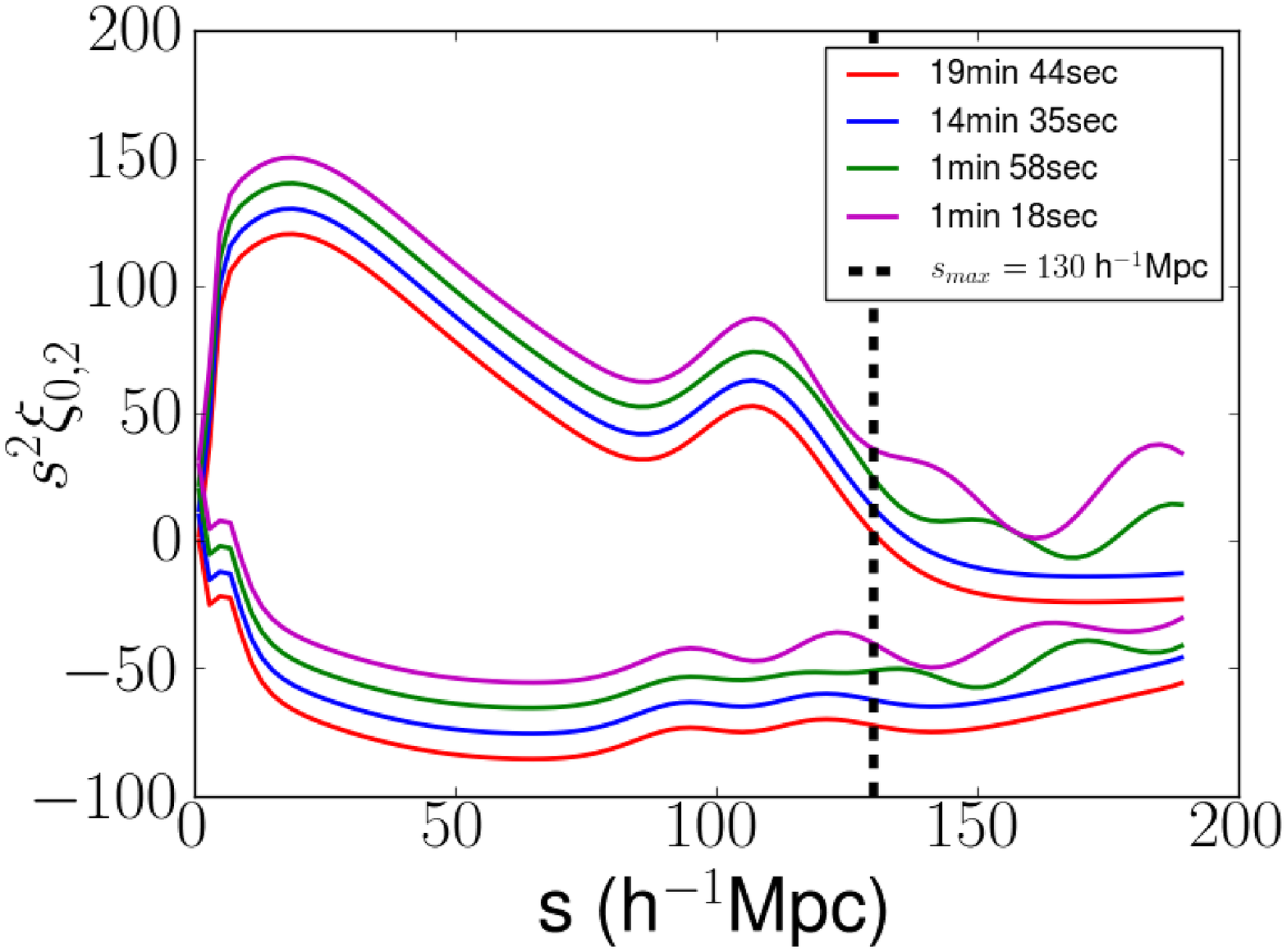}
\includegraphics[width=0.4\textwidth]{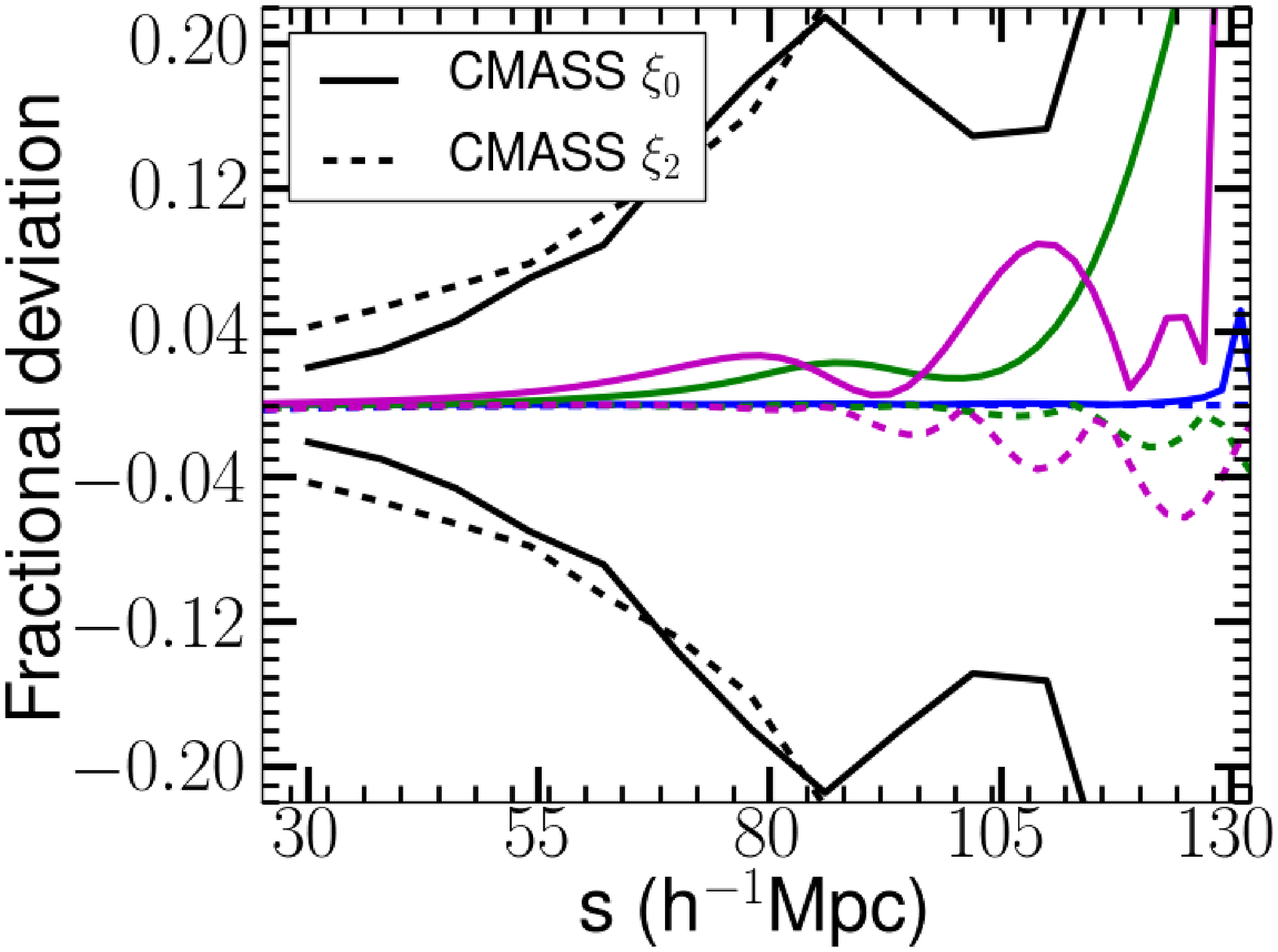}
\caption{Input power spectrum and output correlation function for four different run times . The power spectrum and correlation functions are vertically shifted in order to show the changes clearly. It is clear from the curves that as we sample the power spectrum more sparsely the correlation function does not converge at large scale. This result decides our maximum fitting scale as $126$ h$^{-1}$Mpc for this analysis. The lower panel shows the fractional deviation of correlation function from the completely converged (red line) correlation function. The solid lines represents  monopole and the dashed lines quadrupole. The black line demonstrates the fractional error in our measurement. The error in theoretical model is much smaller than the measurement error. }
\label{fig:clpt-opt-pkxi}
\end{figure}

 We start with the finely-sampled power spectrum and $k_{max}>100$ Mpc$^{-1}$h. The initial power spectrum (Figure \ref{fig:clpt-opt-pkxi} shown in red) has linear sampling for $k<0.5$ with 1030 points. The sampling is logarithmic for $k>0.5$ with 500 points until $k_{max}=100$. The CLPT takes  20 minutes to run with this sampling of power spectrum. The correlation function converges until $r=200$ h$^{-1}$Mpc with the above mentioned sampling. We start with this power spectrum and attempt to resample it in such a way that the run time decreases without introducing any error within the scale of interest which is assumed to be $r<130$ h$^{-1}$Mpc. The results of this optimization are summarized in Figure ~\ref{fig:clpt-opt-pkxi}.

Figure ~\ref{fig:clpt-opt-pkxi}  shows the power spectra, which are vertically shifted for clarity. As we reduce the sampling of linear matter power spectrum, the computation time (shown in the legend of correlation function panel) decreases. As we reduce the sampling in $k$ the correlation function fails to converge for large scale. The fastest runtime (1min 18 seconds) produces a correlation function that doesn't converge above 130 h$^{-1}$Mpc. The corresponding sampling of linear matter power spectrum is as follows (shown in purple in Figure \ref{fig:clpt-opt-pkxi}):

\begin{align}
 \mathbf{k<0.5:}  & \textrm{ linear sampling, 150 points}   \nonumber \\
\mathbf{0.5<k<1.8:}  & \textrm{ logarithmic sampling, 90 points}  \nonumber \\
\mathbf{1.8<k<10:}  & \textrm{ logarithmic sampling, 90 points}  \nonumber
\end{align}

\subsection{Discretizing a subspace of the full parameter space} 
We wish to perform the likelihood minimization on the full parameter space of Planck and RSD parameters. This approach creates a nine-dimensional parameter space, which requires millions of likelihood evaluations; hence, the perturbation theory best case runtime of 1 minute is still too long to achieve convergence in a reasonable time with feasible computing resources.The CLPT, however,  depends only on the input linear power spectrum which is the function of four cosmological parameters \{$\Omega_b h^2$, $\Omega_c h^2$, $n_s$, $H_0$\}. Therefore, we can significantly reduce the number of evaluations of the CLPT correlation function if it is done on the subspace of the full nine-dimensional parameter space. In order to avoid evaluating the CLPT repeatedly for the same cosmology, we have discretized the four-dimensional subspace and run the CLPT calculation only once for each grid in this subspace. Any repetitive call of CLPT for the same grid point will use the stored result from the previous evaluation for that grid point, Significantly increasing the speed of the entire optimization problem. 